\documentclass [prl, superscriptaddress, showpacs, twocolumn]{revtex4}
\usepackage {graphicx}
\usepackage {bm}
\usepackage {amsmath}
\usepackage {hyperref}
\def\figw{0.5}

\begin{document}

\title{Temperature-Dependent X-Ray Absorption Spectroscopy of Colossal
Magnetoresistive Perovskites}

\begin{abstract}
The temperature dependence of the O K-edge pre-edge structure in
the x-ray
absorption spectra of the perovskites La$_{1-x}$A$_{x}$MnO$_{3}$, ($A$ = $Ca$, $%
Sr$; $x = 0.3, 0.4$) reveals a correlation between the
disappearance of the splitting in the pre-edge region and the
presence of Jahn-Teller distortions. The different magnitudes of
the distortions for different compounds is proposed to explain
some dissimilarity in the line shape of the spectra taken above
the Curie temperature.
\end{abstract}

\author{N.~Mannella}
\altaffiliation{Present address: Physics Department, Stanford
University, Stanford, CA, USA} \email[Corresponding author,
electronic address: ]{NMannella@lbl.gov} \affiliation{Department
of Physics, UC Davis, Davis, CA, 95616, USA}
\affiliation{Materials Sciences Division, Lawrence Berkeley
National Laboratory, Berkeley, CA, 94720, USA}
\author{A.~Rosenhahn}
\altaffiliation{Present address: Heidelberg University,
Heidelberg, Germany} \affiliation{Materials Sciences Division,
Lawrence Berkeley National Laboratory, Berkeley, CA, 94720, USA}
\author{M.~Watanabe}
\affiliation{Materials Sciences Division, Lawrence Berkeley
National Laboratory, Berkeley, CA, 94720, USA} \affiliation{RIKEN,
HYOGO 679-5148, Japan}
\author{B.~C.~Sell}
\affiliation{Department of Physics, UC Davis, Davis, CA, 95616,
USA} \affiliation{Materials Sciences Division, Lawrence Berkeley
National Laboratory, Berkeley, CA, 94720, USA}
\author{A~Nambu}
\altaffiliation{Present address: KEK-PF, Tsukuba, Ibaraki,
305-0801, Japan} \affiliation{Materials Sciences Division,
Lawrence Berkeley National Laboratory, Berkeley, CA, 94720, USA}
\affiliation{Department of Chemistry,The University of Tokyo,
Tokyo,113-0033, Japan}
\author{S.~Ritchey}
\affiliation{Department of Physics, UC Davis, Davis, CA, 95616,
USA} \affiliation{Materials Sciences Division, Lawrence Berkeley
National Laboratory, Berkeley, CA, 94720, USA}
\author{E.~Arenholz}
\affiliation{Advanced Light Source, Lawrence Berkeley National
Laboratory, Berkeley CA, 94720, USA}
\author{A.~Young}
\affiliation{Advanced Light Source, Lawrence Berkeley National
Laboratory, Berkeley CA, 94720, USA}
\author{Y.~Tomioka}
\affiliation{Correlated Electron Research Center, Tsukuba, Japan }
\author{C.~S.~Fadley}
\affiliation{Department of Physics, UC Davis, Davis, CA, 95616,
USA} \affiliation{Materials Sciences Division, Lawrence Berkeley
National Laboratory, Berkeley, CA, 94720, USA}

\date{\today}

\maketitle

\section{Introduction}
The discovery of the colossal magnetoresistance (CMR) effect,
i.e., the extremely large drop in resistivity accompanying the
application of a magnetic field near the Curie temperature
(T$_{C}$), has motivated intensive
studies of manganese oxides based upon the perovskite structure \cite%
{Book:1, Coey:99, Ramirez:99,Salamon:01} with the formula unit
ABMnO$_{3}$, (A = rare earth atom, B = divalent atom). Despite
intensive investigations, the CMR oxides are still not fully
understood, thus posing continuing challenges for both experiment
and theory. In fact, it has been recognized for a decade that the
double-exchange (DE) model \cite{Zener:51} which provides a
qualitatively correct description of the CMR effect, needs to be
supplemented by more complex models, not necessarily mutually
exclusive with each other, in order to explain the rich variety of
properties exhibited by the CMR oxides. In particular, the
presence of short-range-ordered Jahn-Teller distortions (JTDs) of
the MnO$_{6}$ octahedra and polaron formation has been intensively
investigated both theoretically and experimentally due to its
importance in the CMR effect \cite{Millis:95, Millis:96, Tyson:96,
Billinge:96, De Teresa:97, Booth:98, Louca:99}.

X-ray absorption spectroscopy (XAS) at the O K-edge has proven to
be a powerful tool for addressing important questions about the
physics of manganites. The O K-edge absorption process is
associated with the O 1s $\longrightarrow \ $Mn 2p dipole
transitions. A typical XAS spectrum of La$_{1-x}$A$_{x}$MnO$_{3}$
(A = Ca, Sr) consists of three main structures which have been
identified as originating from strong hybridization of the O 2p
orbitals with various unoccupied orbitals: Mn 3d at $\approx $ 530
eV (often referred to as the ``pre-edge''
structure), La 5d/4f and Ca 3d/Sr 4d at $\approx $ 535 eV, and Mn 4sp at $%
\approx $ 545 eV \cite{Abbate:92, Abbate:99}. The pre-edge
spectral region extending approximately 4 eV below the primary
absorption threshold and extending over $\approx $ 528--532 eV,
has received a great deal of attention since it represents the Mn
3d-derived unoccupied states. The presence of strong absorption in
the pre-edge region reveals strong hybridization between the O 2p
and Mn 3d states, thus indicating that the holes introduced upon
doping with a divalent metal have a mixed Mn 3d -- O 2p character,
as first recognized by Abbate \cite{Abbate:92} and subsequently by
other authors \cite{Saitoh:95, Pellegrin:97, Park:98, Dessau:01}.

In light of these considerations, XAS spectra of the O K-edge
should reveal signatures for the presence of JTDs, since the
latter are expected to split the degeneracy of the Mn 3d-derived
e$_{g}$ and t$_{2g}$ unoccupied levels. Indeed, the portion of the
pre-edge structure extending from $\approx $ 528.5 eV to $\approx
$ 530.3 eV can exhibit a double peak whose presence/disappearance
has been discussed by some authors as a signature for JTDs. In
particular, Dessau proposed that the presence of the splitting is
a signature of the broken degeneracy of the e$_{g}$ states due to
JTDs of the MnO$_{6}$ octahedra \cite{Dessau:01}. On the other
hand, in a temperature-dependent
study of O K-edge pre-edge structure in Pr$_{0.7}$Sr$_{0.3}$MnO$_{3}$ and Pr$%
_{0.7}$Ca$_{0.15}$Sr$_{0.15}$MnO$_{3}$ perovskites, Toulemonde and
co-workers pointed out the existence of a correlation between the
appearance of the splitting of the O K-edge pre-edge and the
transition from an insulating to metallic state
\cite{Toulemonde:99}. On the contrary, no temperature dependence
of the O pre-edge peak was detected in compounds which show the
same insulating behavior over the whole temperature range, even
when transitions from paramagnetic to ferromagnetic states are
present. The authors thus inferred that the disappearance of the
splitting is a signature for JTDs in the MnO$_{6}$ octahedra.

Due to the significance of the JTDs in the CMR effect, it is thus
important to get more insight into the modifications of the
unoccupied states induced by the presence of JTDs as possibly
revealed by signatures in XAS spectra. In light of this, we have
studied the temperature-dependence of the O K-edge
pre-edge structure in La$_{0.7}$Ca$_{0.3}$MnO$_{3}$ (LCMO) and La$_{1-x}$Sr$%
_{x}$MnO$_{3}$ (LSMO, x = 0.3, 0.4) perovskites, both of which
compounds
which have been shown to be characterized by the presence of JTDs above T$%
_{C}$ \cite{Tyson:96, Billinge:96, Booth:98, Louca:99, io,
Shibata:03}. Our data show a correlation between the absence of
the splitting in the pre-edge region and the presence of JTDs,
even when no metal-insulator transition takes place and the
electronic phase is metallic for all the temperatures accessed by
the experiments. Minor differences exhibited in the
high-temperature spectra of LCMO and LSMO are explained in terms
of the relative magnitude of the JTDs.

\section{Experimental}
We used high-quality single crystals grown by the floating-zone
method, with details on growth conditions and samples
characterization being reported elsewhere \cite{Urushibara:95}.
The compounds studied exhibit a ferromagnetic-to-paramagnetic
phase transition at T$_{C}$ $\approx $ 250 K
(LCMO, x = 0.3), T$_{C}$ $\approx $ 365 K (LSMO, x = 0.3), and T$_{C}$ $%
\approx $ 370 K (LSMO, x = 0.4) \cite{Urushibara:95, Schiffer}.
While LCMO undergoes a metal-insulator transition as the
temperature is raised above T$_{C}$ \cite{Schiffer}, LSMO is
always in a metallic state for the compositions studied
\cite{Urushibara:95}. The crystals have been fractured at room
temperature in situ at base pressures better than 2 $\times $
10$^{-10}$ torr. Quantitative x-ray photoelectron spectroscopy
analyses of core-level peaks have routinely confirmed the expected
stoichiometries and doping levels x to within experimental
accuracy ($\Delta $x $\approx $ $\pm $0.03), and further shown
that the degree of surface stoichiometry alteration or
contamination was negligible, indicating a high surface
cleanliness and stability.

\begin{figure}[htbp]
\centerline{\includegraphics[width=\figw\textwidth]{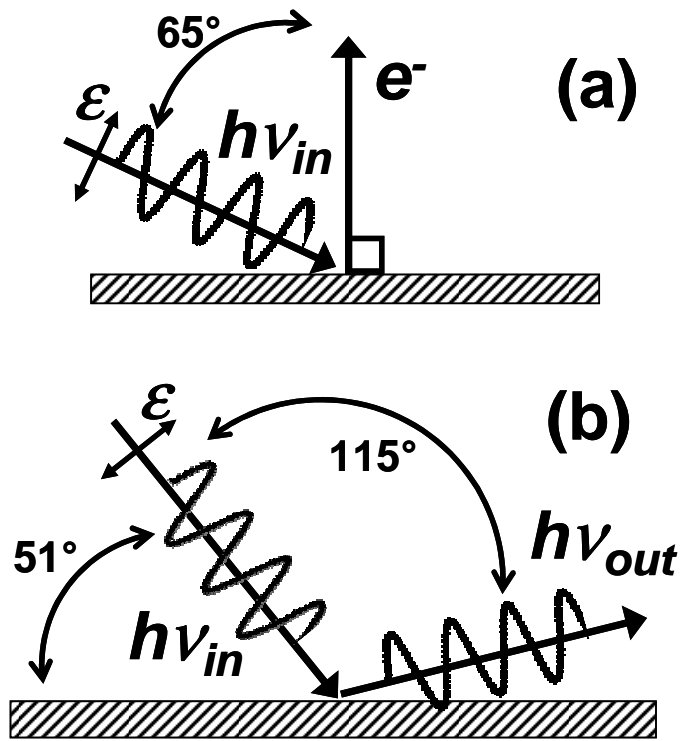}}
\caption{ Schematic layout of the experimental geometries for the
electron yield (a) and fluorescence yield modes (b),
respectively\label{fig:1} }
\end{figure}

The data have been taken with the Multi-Technique
Spectrometer/Diffractometer endstation located on the
elliptically-polarized undulator beamline 4.0.2 at the Berkeley
Advanced Light Source, which permits photoemission, x-ray
emission, and x-ray absorption to be carried out on a single
sample. The x-ray absorption spectra reported here have been
excited with linearly p-polarized light and detected with
secondary electrons of $\approx $ 100 eV kinetic energy collected
with a Scienta ES200 electron analyzer and fully corrected for
detector non-linearity effects according to a procedure described
elsewhere \cite{Nonlin}. In order to maximize the probing depth,
the spectra were collected in normal emission, i.e. with the
electron analyzer set perpendicular with respect to the surface.
Additional spectra were collected in the more bulk sensitive
fluorescent yield mode by integrating the x-ray emission signal of
the O 2p $\rightarrow $ O 1s transition using a Scienta XES 300
grating spectrometer. Additional information regarding the
experimental geometry is shown in Fig. 1. The instrumental
resolution, as controlled by the beamline monochromator, was set
to 150 meV.

\section{Results And Discussion}

The O K pre-edge regions as measured with secondary electrons on La$_{0.7}$Ca%
$_{0.3}$MnO$_{3}$, La$_{0.6}$Sr$_{0.4}$MnO$_{3}$ and La$_{0.7}$Sr$_{0.3}$MnO$%
_{3}$ at several temperatures are shown in Figs. 2, 3a and 3b,
respectively. Two main structures are observed in the pre-edge
spectra, one extending from $\approx $ 528.5 to 530.3 eV and which
can exhibit either one broad peak or two separated peaks, and a
weaker one centered at $\approx $ 531.5 eV. The
relevant Mn 3d states involved in the pre-edge region are majority spin e$%
_{g}$ (e$_{g}^{\uparrow }$), minority spin t$_{2g}$
(t$_{2g}^{\downarrow }$), and minority spin e$_{g}$
(e$_{g}^{\downarrow }$). Many authors assign the first
structure over $\approx $ 528.5 to 530.3 eV to O hybridization with e$%
_{g}^{\uparrow }$ and t$_{2g}^{\downarrow }$ states, while the
second structure, located $\approx $ 2 eV higher in energy at
531.5 eV, has been assigned to e$_{g}^{\downarrow }$ states
\cite{Abbate:92, Abbate:99, Pellegrin:97, Park:98, Dessau:01,
Toulemonde:99}.

The most remarkable characteristic of our data is the temperature
dependence of the first structure located at 528.5 -- 530.3 eV. In
particular, this structure consists of two distinct peaks
separated by $\approx $ 0.65 eV for temperatures lower than
T$_{C}$ (cf. Fig. 2 and Fig. 3), while the splitting disappears
for temperatures above T$_{C}$ (cf. Fig. 2, 3b and 3c). Also
noticeable is the suppression of the structure at 531.5 eV for
temperatures higher than T$_{C}$, as most clearly shown in the
spectra collected in the fluorescent yield mode (cf. Fig. 3c). We
can rule out the possibility that this temperature dependent
behavior could be due to surface effects, since
the spectra collected in the more bulk sensitive fluorescent yield mode in La%
$_{0.7}$Sr$_{0.3}$MnO$_{3}$ exhibit the same spectral features
(Fig. 3c), with the only exception being that the structure at
$\approx $ 531.5 eV is much more pronounced, possibly due to the
different detection methods resulting in different matrix element
and fluorescent-yield effects. This structure is equally well
prominent in the fluorescent yield spectra reported in ref.
\cite{Toulemonde:99}.

\begin{figure}[htbp]
\centerline{\includegraphics[width=\figw\textwidth]{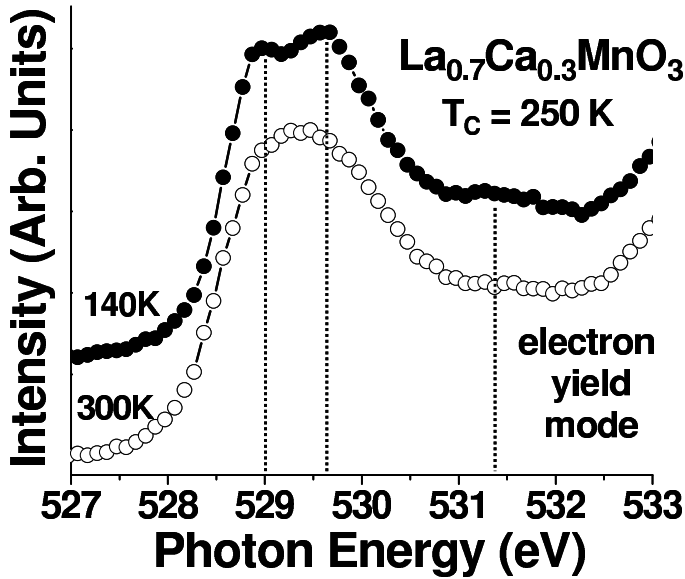}}
\caption{ Temperature dependence of the pre-edge structure in O
K-edge XAS spectra from La$_{0.7}$Ca$_{0.3}$MnO$_{3}$. The
splitting visible at 140 K disappears for temperatures above
T$_{C}$, resulting in a broad peak consistent with the unfolding
of the energy levels induced by the Jahn-Teller distortions
(JTDs).\label{fig:2} }
\end{figure}

The assignment of the splitting in the main pre-edge structure is
still a matter of controversy in the literature. Pellegrin
\cite{Pellegrin:97} and Park \cite{Park:98} assigned the first of
the two peaks (529 eV) to e$_{g}^{\uparrow }$ states, the second
one (529.65 eV) to the t$_{2g}^{\downarrow }$ states. Dessau
assigned both peaks to e$_{g}^{\uparrow }$ states after observing
that, although expected in the same energy range, the
t$_{2g}^{\downarrow }$ states should have a low intensity because
of the weak hybridization between the O 2p and Mn t$_{2g}$ states,
as predicted in a cluster model calculation by Saitoh
\cite{Saitoh:95, Dessau:01}. Consequently, Dessau interpreted the
splitting as a signature of the broken degeneracy of the e$_{g}$
states due to a JTD of the MnO$_{6}$ octahedra. On the contrary,
Toulemonde and co-workers observed that the splitting appears when
a transition from a paramagnetic-insulating to a
ferromagnetic-metallic state occurs, thus arguing that the
presence of strong JTDs is revealed by the disappearance of the
splitting \cite{Toulemonde:99}. To interpret their data,
Toulemonde et al. performed a crystal field analysis based on
cluster calculations reported by Kurata and Colliex \cite{Kurata}
and supplemented by introducing the splitting of the e$_{g}$ and
t$_{2g}$ molecular orbitals due to distortions of the manganese
octahedra \cite{Toulemonde:99}. Such strong distortions would be
expected to split the degeneracy of both the e$_{g}$ and t$_{2g}$
levels, resulting in a set of energy states whose separation is
smaller than the experimental resolution, and leading to a broad
peak in the pre-edge structure of lowest energy. In this second
interpretation, for compounds which exhibit an insulator-metal
transition on lowering the temperature below T$_{C}$, the
reduction of the JTD reduces considerably the unfolding of the
e$_{g}$ and t$_{2g}$ states, resulting in a clear separation of
the t$_{2g}^{\downarrow }$ from the e$_{g}^{\uparrow }$ states,
visible in the dominant pre-edge structure as the presence of two
distinct peaks \cite{Toulemonde:99}.

Turning now to our experimental data for LCMO (Fig. 2), a compound
generally agreed to be characterized by local JTDs \cite{Tyson:96,
Billinge:96, Booth:98}, there is full agreement
with what is reported by Toulemonde on Pr$_{0.7}$Sr$_{0.3}$MnO$_{3}$ and Pr$%
_{0.7}$Ca$_{0.15}$Sr$_{0.15}$MnO$_{3}$ perovskites. That is, at
room temperature, which is above the temperature at which both a
ferromagnetic-to-paramagnetic and a metal-insulator transition
take place, the double peak structure observed below T$_{C}$
disappears and JTDs are fully developed \cite{Tyson:96,
Billinge:96, Booth:98}, resulting in a broad peak centered at
$\approx $ 529.4 eV. These results seem to rule out the hypothesis
proposed by Dessau according to which the splitting is a signature
for JTDs, since the splitting is clearly visible at 140K where the
JTDs are weak and disappear at high temperatures (300K) where the
JTDs are fully developed \cite{Tyson:96, Billinge:96, Booth:98}.

For La$_{0.6}$Sr$_{0.4}$MnO$_{3}$ (Fig. 3a), no substantial change
in the splitting is observed when both spectra are measured below
T$_{C}$, in the metallic state, in full agreement with what is
reported by Toulemonde.

However, the existence of a correlation between the disappearance
of the splitting and the transition from a metallic to an
insulating state as
proposed by Toulemonde is less obvious in the case of La$_{0.7}$Sr$_{0.3}$MnO%
$_{3}$ (Fig. 3b and 3c): in fact, while the double peak in the
pre-edge structure is clearly visible below T$_{C}$, it disappears
above T$_{C}$, even though La$_{0.7}$Sr$_{0.3}$MnO$_{3}$ is a
metal over the whole temperature range accessed by our
measurements. Our results are however in complete agreement with
the La$_{0.7}$Sr$_{0.3}$MnO$_{3}$ data reported by Park, who
interpreted the disappearance of the splitting as a spectral
weight transfer associated with the reduction of the density of the e$%
_{g}^{\uparrow }$ states at the Fermi level at high temperatures
\cite{Park:98}.

Although the temperature dependence of the pre-edge peak splitting
is qualitatively similar for LCMO and LSMO, the line shapes above
T$_{C}$ show some dissimilarity. Indeed, for both compounds the
spectra show some broadening above T$_{C}$ for photon energies $>$
$\approx $ 528.75, but while in LCMO the disappearance of the
splitting above T$_{C}$ results in a broad peak centered at
$\approx $ 529.4 eV, for LSMO it seems to take place primarily
through a loss of intensity of the peak located at 529 eV, thus
producing a broad peak less symmetric than in LCMO.

\begin{figure}[htbp]
\centerline{\includegraphics[width=\figw\textwidth]{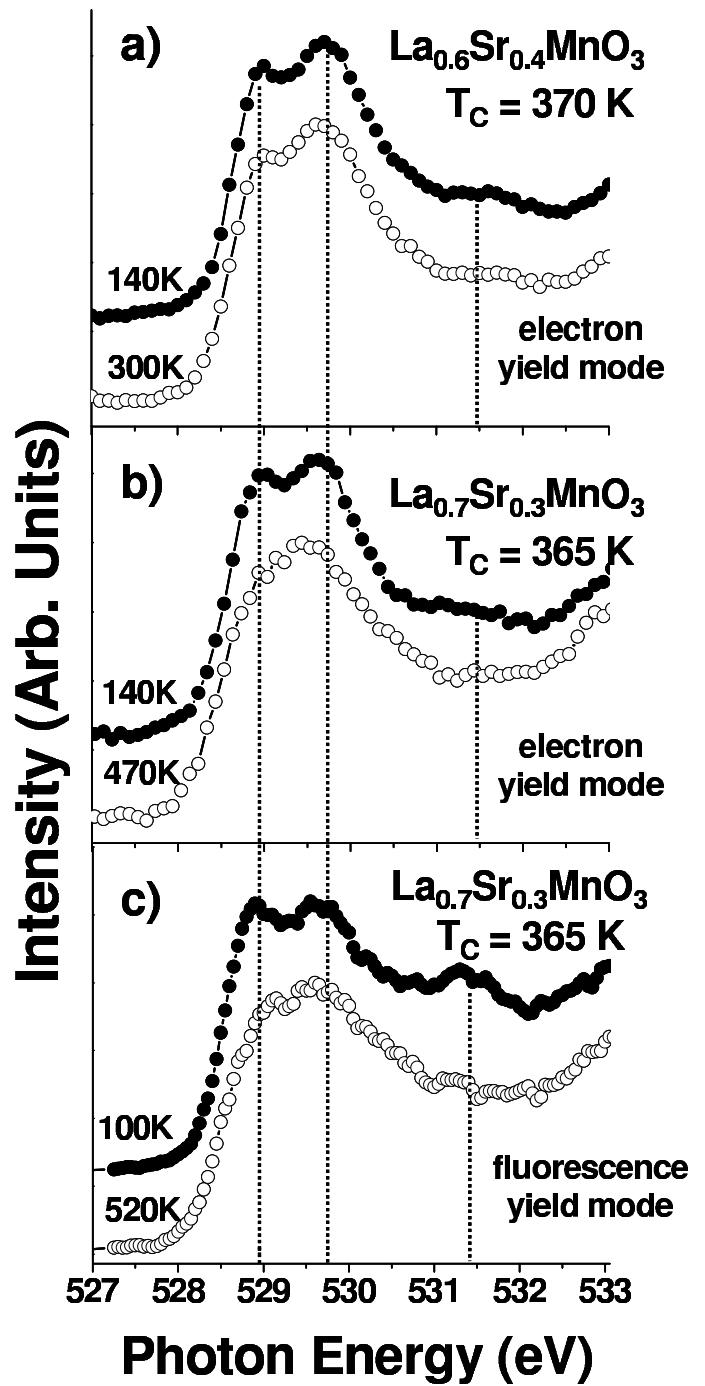}}
\caption{ Temperature dependence of the pre-edge structure in O
K-edge XAS
spectra from La$_{0.6}$Sr$_{0.4}$MnO$_{3}$ (a) and La$_{0.7}$Sr$_{0.3}$MnO$%
_{3}$ (b and c). The spectra have been taken both in the electron
yield (a-b) and fluorescent yield modes (c). The splitting is
clearly visible in both compositions for temperatures below
T$_{C}$. Above T$_{C}$ (b-c), the splitting disappears, primarily
due to the loss of intensity of the structure at 529 eV, resulting
in a single peak less broadened than in LCMO.\label{fig:3} }
\end{figure}

At first, it may thus seem like the results reported in Figs. 3b
and 3c cannot be explained by the model proposed by Toulemonde.
Nonetheless, we point out that in a recent multi-spectroscopic
study of LSMO (x = 0.3, 0.4),
we have observed an increase in the average spin moment of the Mn atom from $%
\approx $ 3 to $\approx $ 4 $\mu _{B}$, corresponding to about 1
electron transferred to the Mn atom. This increase was concomitant
with local JTDs at high temperature, providing strong evidence for
direct detection of lattice polaron formation in the paramagnetic
metallic phase \cite{io}.

In light of the results revealed by our previous work, we
interpret the loss of intensity of the peak located at 529 eV and
the consequent disappearance of the splitting as signatures for
Jahn-Teller (JT) polaron formation. In fact, the loss of intensity
of the peak located at 529 eV can be easily explained by a
reduction of the density of the lowest unoccupied states due to
the electron localization above T$_{C}$ associated with JTDs, and
resulting reduced O 2p-Mn 3d hybridization, an interpretation
consistent with the one provided by Park \cite{Park:98}.

Our results on LSMO and the interpretation framework proposed by
Toulemonde are in addition not inconsistent once we acknowledge
that polarons and metallic conductions are not mutually exclusive.
We suggest that the differences in the line shapes above T$_{C}$
for LCMO and LSMO are due to the difference in the magnitude of
the JTDs for the two compounds. In fact, recent EXAFS results
indicate that the size of the JTDs in LSMO is only about half that
in the Ca-doped compound \cite{io}. It is thus possible that in
LSMO the reduced JTDs are not as effective as for LCMO in lifting
the degeneracy of the t$_{2g}^{\downarrow }$ and e$_{g}^{\uparrow
}$ states, which remain separated even when the JTDs are fully
developed at high temperature, with the result that the splitting
disappears by evolving into a structure less broadened than in
LCMO \cite{magni}.

Overall, our data thus seem to support the scheme previously
proposed by Toulemonde and further indicate that the correlation
between the absence of the splitting in the pre-edge region and
the presence of JTDs may occur even when the electronic phase is
metallic, without requiring a metal-insulator transition.
According to Toulemonde, the lowest unoccupied state, as
determined by the degree of distortion and by the relative
magnitude between
the exchange interaction and crystal field strength, is a t$%
_{2g}^{\downarrow }$ state with filling depending on the
substitution amount \cite{Toulemonde:99}. However, on the basis of
the data reported, we can only assert that our results are
qualitatively consistent with the scenario proposed by Toulemonde,
but we cannot address the assignment of the peaks at 529 and
$\approx $ 529.7 eV, an issue which requires further
investigations.

\section{Conclusions}
In conclusion, we have measured the temperature dependence of the
O K-edge pre-edge structure in La$_{1-x}$A$_{x}$MnO$_{3}$, (A =
Ca, Sr, x = 0.3, 0.4) perovskites. Our data are in qualitative
agreement with an interpretation scheme previously proposed
according to which there exists a correlation between the
disappearance of the splitting in the pre-edge region and the
presence of JTDs. The difference in the magnitude of the JTDs for
LSMO and LCMO likely explains the dissimilar line shape of the
high temperature spectra.

\section{Acknowledgments}
We thank M. West and A. Mei for assistance with the measurements;
Z. Hussain for instrumentation development; Y. Tokura for
assistance with obtaining samples. A. Rosenhahn gratefully
acknowledges the support of the Alexander von Humboldt foundation
by a Feodor-Lynen research grant. This work was supported by the
Director, Office of Science, Office of Basic Energy Sciences,
Materials Science and Engineering Division, U.S. Department of
Energy under Contract No. DE-AC03-76SF00098.


\begin{thebibliography}{99}
\bibitem{magni} The different magnitude of the JTDs for LSMO and LCMO would
qualitatively correspond to the cases respectively illustrated in
Fig. 8b and Fig. 8c of ref. [19].
\end{thebibliography}
\end{document}